\documentclass[
reprint,
superscriptaddress,
amsmath,amssymb,
aps,
]{revtex4-2}

\usepackage{graphicx}
\usepackage{dcolumn}
\usepackage{bm}
\usepackage{kotex}
\usepackage{svg}
\usepackage{hyperref}
\hypersetup{
    colorlinks=true,
    linkcolor=blue,
    citecolor=blue,
}

\begin{document}


\title{Scale-freeness under node removal: a finite-size scaling perspective}

\author{Yeonsu Jeong}
\affiliation{Department of Applied Physics, Hanyang University, Ansan 15588, South Korea}

\author{Deok-Sun Lee}
\email{deoksunlee@kias.re.kr}
\affiliation{School of Computational Science, Korea Institute for Advanced Study, Seoul 02455, South Korea}

\author{Mi Jin Lee}
\email{mijinlee@pusan.ac.kr}
\affiliation{Department of Physics, Pusan National University, Busan 46241, South Korea}

\author{Seung-Woo Son}%
\email{sonswoo@hanyang.ac.kr} 
\affiliation{Department of Applied Physics, Hanyang University, Ansan 15588, South Korea}

\date{\today}

\begin{abstract}
In heterogeneous network systems such as ecological and social networks, structural stability depends on how connectivity changes under node removal, as different removal sequences can trigger distinct modes of systemic collapse. While robustness to random failures and targeted attacks has been extensively studied, most analyses have focused on connectivity loss or degree distribution, rather than on how scale-invariant organization emerges and evolves with system size. Here we examine how scale-free structure evolves under progressive degree-dependent node removal, systematically varying the hub-protection strength $\theta$. Starting from scale-free networks, we apply the recently developed finite-size scaling (FSS) analysis to node-removed networks and compare the results with those from Kullback-Leibler (KL) divergence-based classification. We find that under random ($\theta=0$) and hub-protecting removal ($\theta>0$), the two criteria largely agree, whereas under hub-preferential removal ($\theta<0$), networks may appear scale-free according to the KL criterion while failing the FSS test of scaling collapse. This discrepancy indicates that similarity to a reference degree distribution does not guarantee the persistence of scale-invariant organization across system sizes. The two diagnostics thus probe complementary aspects of network structure, and their joint use provides a more complete characterization of structural degradation. 


\end{abstract}

\maketitle

\section{Introduction}
The structure of connectivity plays a fundamental role in complex systems, strongly influencing dynamical behaviors such as spreading processes, resilience, and robustness against perturbations. Since the discovery of scale-free (SF) networks~\cite{barabasi1999sf}, the presence of hub nodes has been recognized as structurally and dynamically consequential. In epidemic spreading, for example, hub nodes can simultaneously accelerate transmission and serve as effective intervention targets~\cite{pastor2001epidemic,lee2019epidemic}. Consequently, assessing the degree of connectivity heterogeneity has become a central task in network science~\cite{clauset2009power,broido2019scale,voitalov2019scale}.

Many studies of scale-free organization have focused on growing networks, where preferential attachment mechanisms generate power-law degree distributions. In contrast, networks may undergo structural degradation through node failures or targeted removal~\cite{albert2000error,cohen2000resilience}. The robustness of such networks has been extensively investigated, primarily in terms of connectivity loss, percolation thresholds, or epidemic spreading dynamics. At the same time, a substantial body of work has examined how sampling or incomplete observation affects the inference of power-law degree distributions~\cite{stumpf2005subnets,clauset2009power,broido2019scale,lee2006snowball,son2012directed}. Unlike sampling problems, in which the underlying structure remains intact but is only partially observed, the progressive reduction of a network through node removal alters the network structure itself. Under such structural evolution, a natural question is how structural signatures associated with scale-freeness evolve as the network progressively degrades. In addition, degree-based descriptions do not directly capture global connectivity, which is closely related to network functionality, and changes in degree statistics do not necessarily imply corresponding changes in network-level connectivity, such as those characterized by the giant connected component.

Recent work has investigated degree-dependent node-removal strategies to model different scenarios of network collapse and to understand how scale-freeness responds to such contraction~\cite{lee2022degree}. It was shown that hub-preferential removal, corresponding to targeted attacks~\cite{albert2000error}, induces a crossover from an SF-like structure to an Erd\H{o}s-R\'enyi (ER)-like structure as nodes are progressively removed, whereas the SF-like regime was reported to persist under random failure and hub-protecting removal. In that study, structural regimes were classified as SF-like or ER-like using the relative entropy, namely the Kullback-Leibler (KL) divergence~\cite{KL1951entropy}, with respect to reference degree distributions~\cite{tishby2019convergence,tishby2020analysis,budnick2022structure,lee2022degree}. This entropy-based perspective provides a quantitative way to track changes in degree statistics during contraction. However, similarity to prescribed degree distributions does not necessarily imply that scale-invariant organization remains stable across system sizes.

To complement this perspective, we employ the finite-size scaling (FSS) method proposed in Ref.~\cite{serafino2021true}. In statistical physics, FSS is a standard framework for identifying genuine critical behavior by examining scaling collapse across different system sizes~\cite{privman1990fss}. In network science, FSS has also been widely used to analyze critical phenomena such as percolation and epidemic transitions~\cite{dorogovtsev2008critical}. The recent study~\cite{serafino2021true} has developed the FSS as a diagnostic tool to assess scale-freeness beyond the form of a single degree distribution, by directly testing whether scaling behavior is maintained across scales.

In this work, we investigate how scale-freeness evolves under degree-dependent network contraction by combining entropy-based, scaling-based, and connectivity-based diagnostics. By systematically comparing the KL divergence and FSS across different removal fractions and hub-protection levels, we examine the conditions under which SF-like degree statistics are maintained or break down as the network collapses, and identify regimes in which the two diagnostics yield consistent or inconsistent classifications. While such classifications provide a useful summary of structural tendencies, our objective is not limited to a binary identification of scale-freeness, but rather to develop a more nuanced understanding of network structure by jointly examining multiple complementary diagnostics. In addition, by analyzing the giant connected component (GCC), we examine how global connectivity evolves alongside these structural signatures. Each diagnostic probes a distinct aspect of network structure, with KL divergence capturing similarity to reference distributions, FSS evaluating the stability of scaling behavior across system sizes, and the GCC reflecting global connectivity. This combined approach provides a more comprehensive view of network structure and highlights the need for a multi-perspective understanding of structural organization under node removal.

The remainder of this paper is organized as follows. In Sec.~\ref{sec:model_methods}, we introduce the degree-dependent node-removal process and describe the FSS framework used to assess scale-freeness. Section~\ref{sec:results} presents the results of the FSS and KL-divergence analyses under network contraction. In Sec.~\ref{sec:discussion}, we discuss the implications of these findings for structural degradation and dynamical vulnerability in complex networks.

\section{Model and Methods}
\label{sec:model_methods}
\subsection{Degree-based node removal process} 
\label{subsec:node_removal}
The node-removal strategy depends on the scenario being modeled. Different removal strategies (e.g., random failure or targeted attack) affect the functionality of networked systems in distinct ways, resulting in different levels of robustness or resilience against perturbations. Consequently, even when nodes are removed or sampled based on the same level of centrality, the resulting structural changes can differ depending on the underlying connectivity~\cite{posfai2016network}.

As a concrete implementation of such scenario-dependent removal strategies, recent work~\cite{lee2022degree} introduced a degree-based node-removal probability for a node with degree $k$ at time $\tau$ as
\begin{align} \label{eq:qk}
q_k(\tau) = \frac{(k+1)^{-\theta}}{\sum_{i\in I_{\rm surv}(\tau)}{(k_{i}(\tau)+1})^{-\theta}},
\end{align}
at every discrete time steps $\tau=0, 1, 2, \cdots$, where $I_{\rm surv}(\tau)$ is the set of surviving nodes at $\tau$ and $k_i(\tau)$ is the degree of node $i$ in the induced subgraph. The removal probability is controlled by the hub-protection level $\theta$. Representatively, $\theta=-1, 0,$ and 1 correspond to hub-preferential ($\theta<0$), random, and hub-protecting ($\theta>0$) removal methods, respectively. The term $(k+1)$, not $k$, is introduced to avoid a singular probability for isolated nodes. In the present study, we adopt this general node-removal process as a baseline framework to explore the emergence of a scale-free regime.

The removal process is implemented as follows. Starting from a network consisting of $N$ nodes and $L$ links, at each time step $\tau$ ($0\leq \tau\leq Nf$), a node is randomly selected and removed with the removal probability $q$ defined in Eq.~(\ref{eq:qk}). Repeating this procedure yields a remaining network of $N_f\equiv N(1-f)$ nodes. By varying the fraction $f$ of removed nodes for a given $\theta$, one can generate networks with different structural properties.

To characterize the structural heterogeneity induced by node removal, Lee et al.~\cite{lee2022degree} analyzed the connectivity of the static-model SF 
network~\cite{goh2001universal} and the Barab\'{a}si--Albert (BA) model~\cite{barabasi1999sf} in the $f$--$\theta$ plane by comparing degree distributions. This was achieved by comparing the KL-
divergence between the empirical degree distribution and the reference degree distributions of the static model and the ER 
random network, defined as
\begin{align}
    S_f^{\rm(PO)}&\equiv\sum_k{p_f(k) \ln{\left[\frac{p_f(k)}{p^{\rm (PO)}(k;m_f)}\right]}},\label{eq:S_PO}\\
    S_f^{\rm(SF)}&\equiv\sum_k{p_f(k) \ln{\left[\frac{p_f(k)}{p^{\rm (SF)}(k;\alpha;m_f)}\right]}},\label{eq:S_SF}
\end{align}
where $p_f(k)$ denotes the empirical degree distribution after removing a fraction $f$ of nodes. The reference distributions $p^{\rm (PO)}(k;m_f)$ and $p^{\rm (SF)}(k;\alpha;m_f)$ with the degree exponent $\alpha$ and mean degree $m_f$ correspond to the analytic forms of the Poisson (PO) distribution~\cite{erdHos1959random} and the static-model SF distribution~\cite{goh2001universal}, respectively.

Within this reference-based framework, the network is classified as being closer to the static-model SF network than to the ER network when $S_f^{\rm(SF)}<S_f^{\rm(PO)}$, and vice versa. Accordingly, the condition $S_f^{\rm(SF)}=S_f^{\rm(PO)}$ defines a boundary $f^*$ separating SF-like and ER-like, or equivalently PO-like, regimes. However, this evaluation scheme has an intrinsic limitation in that the results depend sensitively on the choice of reference distributions and do not provide a direct assessment of scale-freeness itself. To address this issue and identify the intrinsic SF nature in the $f$--$\theta$ plane, we employ the FSS 
method proposed in Ref.~\cite{serafino2021true}, which operates solely on the given network without relying on reference distributions.

\subsection{Finite-size scaling framework for scale-freeness}
\label{subsec:fss}
We illustrate the FSS procedure developed in Ref.~\cite{serafino2021true} in this section. The FSS analysis is applied only when the available scaling range of the degree is sufficiently broad. The minimum degree $k_{\rm min}$ defining the valid range for power-law fitting is estimated by the Kolmogorov--Smirnov (KS) statistic~\cite{Hollander2014Nonparametric} as treated in the maximum likelihood method~\cite{clauset2009power}, and the range $k\geq k_{\rm min}$ is used for further analysis. Let $N^*$ denote the number of nodes involved in this range, i.e., $\int_{k_{\rm min}}^{k_{\rm max}}p(k)dk=N^*/N$, where $k_{\rm max}$ is the empirical maximum value. If $N^*\leq \ln N$, the network is classified as non-scale-free (non-SF), and the FSS procedure is not applied. For networks satisfying $N^*>\ln N$ (hereafter referred to as \emph{feasible}), the FSS analysis is performed.

For feasible networks, we consider the complementary cumulative distribution function (CCDF) $P_f(k)$ after removing a fraction $f$ of nodes, defined as
\begin{equation}
P_f(k)=\int_{k}^{k_{\rm max}}p_f(k^{\prime}) \,dk^{\prime}\sim k^{-\gamma},
\label{eq:ccdf}
\end{equation}
with the CCDF exponent $\gamma=\alpha-1$. We assume that the CCDF oberys the following finite-size scaling:
\begin{equation}
    P_f(k; n_f)=k^{-\gamma}\,g(kn_f^{d}),
    \label{eq:fss_form}
\end{equation}
where $n_f\in\left\{{N_f}/{4}, {N_f}/{2}, {3N_f}/{4}, N_f\right\}$~\cite{serafino2021true} is the  size of subnetworks randomly sampled from the remaining network of size $N_f$, as illustrated in Fig.~\ref{fig:FSSresult}(a). Here, $g(\cdot)$ denotes a scaling function and $d$ is the scaling exponent characterizing the moment ratio $\langle k^i\rangle /\langle k^{i-1}\rangle\sim N^{-d}$ for system size $N$, provided $i-1>\gamma$. A successful scaling collapse across different values of $n_f$ with appropriate exponents $\gamma$ and $d$ indicates the persistence of scale-invariant structure. Note that subnetworks are constructed by random node sampling, which may not preserve mesoscopic structures such as communities or core-periphery organization.

To quantify the quality of collapse of the scaled CCDFs, we consider the ordered degree sequences $k_{n,j}$ of the sampled networks, where $n$ indexes the networks and $j$ denotes the degree rank ($k_{n,j}\leq k_{n,j+1}$), and define the objective function 
\begin{equation}
Q=\frac{1}{3|M|}\sum_{(n,j)\in M}\frac{\left(y_{n,j}-Y_{n,j}\right)^2}{dy_{n,j}^2+dY_{n,j}^2},
\label{eq:Q}
\end{equation}
where $x_{n,j}=k_{n,j}n^d$ and $y_{n,j}=P(k_{n,j})k_{n,j}^{\gamma}$, along with   $dy_{n,j}=dP(k_{n,j})k_{n,j}^{\gamma}$ assuming Poisson errors $dP$, are obtained from empirical data, and $Y_{n,j}$ and $dY_{n,j}$ are the weighted least-squares linear fit to $y_{n,j}$ and its standard error at $x_{n,j}$. The details of Eq.~(\ref{eq:Q}) are described in Ref.~\cite{serafino2021true}. The optimal values of $\gamma$ and $d$ are determined simultaneously by minimizing $Q$ using the Nelder--Mead method~\cite{nelder1965simplex,melchert2009autoscale}.

The scale-freeness of feasible networks is finally assessed according to the magnitude of $Q$ in the spirit of a chi-squared statistic: (i) non-SF for $Q>3$, (ii) weakly SF for $1<Q\leq3$, and (iii) strongly SF for $Q\leq1$~\cite{serafino2021true}. In this work, we adopt a binary classification and regard networks as SF when $Q\leq3$ and non-SF otherwise.

\section{Results}
\label{sec:results}
\subsection{Scale-freeness under degree-based node removal}
\label{subsec:applyFSS}

\begin{figure*}[t]
\includegraphics[width=0.87\linewidth]{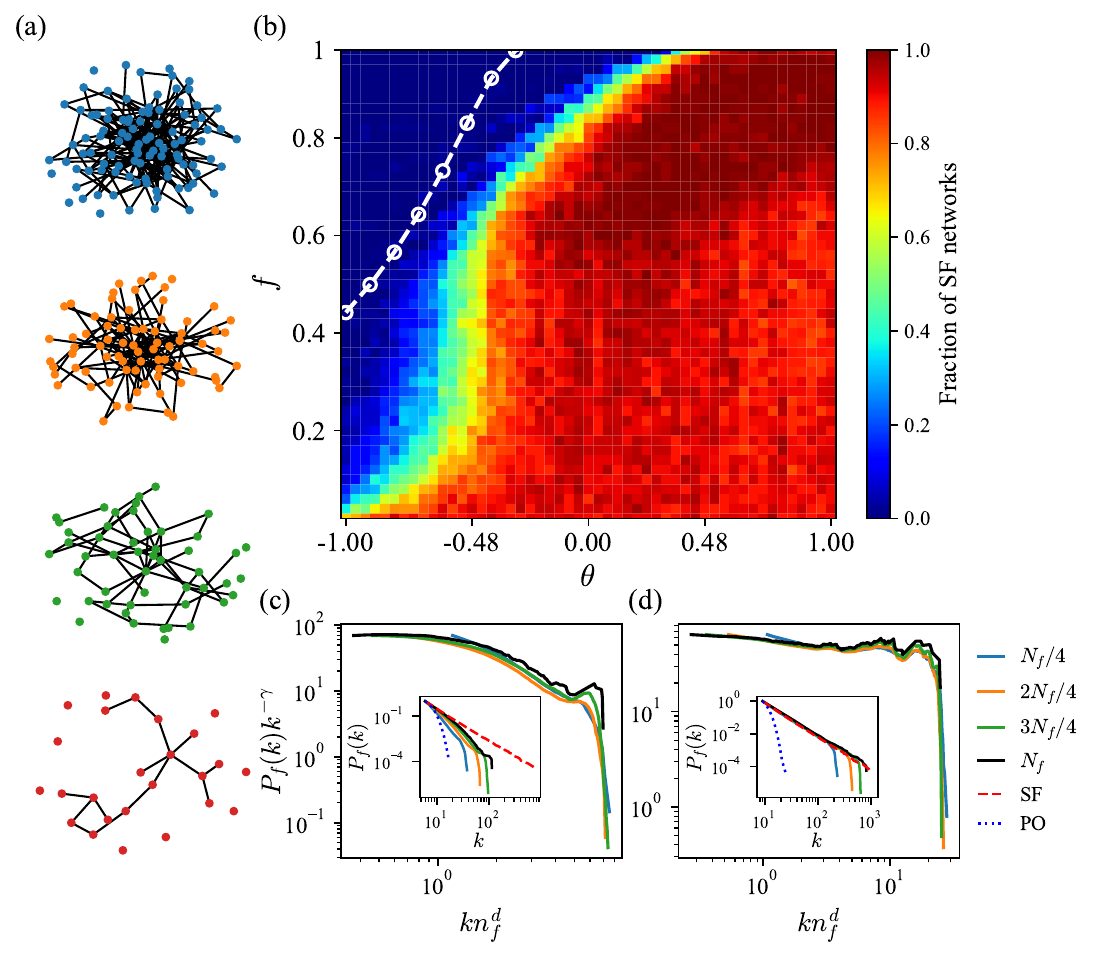}
\caption{FSS tests on BA networks under a general node-removal process. (a) Schematic illustration of the subnetwork sampling procedure used in the FSS analysis. The top network generated by the BA model is regarded as the original network with $N=100$ nodes and $L=257$ links, while the others are randomly sampled subnetworks of sizes ${3N}/{4}$, ${N}/{2}$, and ${N}/{4}$. (b) Fraction of realizations classified as SF-like in the plane of removal fraction $f$ and hub-protection level $\theta$ [see Eq.~(\ref{eq:qk})] for the BA model with $N=N_{(f=0)}=10^5$ and mean degree $\langle k\rangle=10$. The dashed line represents the theoretical boundary $f^*$ between SF-like and PO-like regimes derived from Ref.~\cite{lee2022degree}. (c, d) FSS scaling collapse of the CCDF for $n_f\in\{N_f/4, N_f/2, 3N_f/4, N_f\}$ at (c) $f=0.2$ and $\theta=-1$ yielding $Q=18.2$ and at (d) $f=0.2$ and $\theta=0$ yielding $Q=1.25$. Insets show the original CCDFs with Poisson and scale-free reference distributions.
} 
\label{fig:FSSresult}
\end{figure*}

\begin{figure*}[t]
\includegraphics[width=1\linewidth]{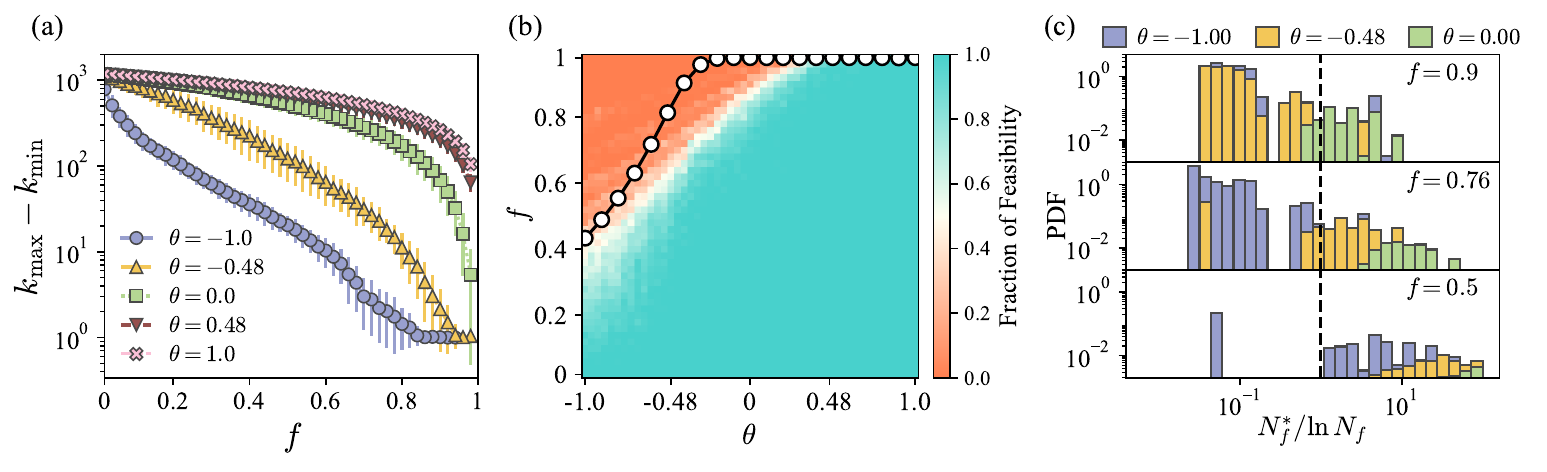}
\caption{Basic statistics of the available scaling region of the degree. (a) Removal fraction $f$ versus $(k_{\rm max}-k_{\rm min})$ for the removal strategy $\theta=-1, -0.48, 0, 0.48$, and $1$, shown representatively. Shaded regions indicate the standard deviations. (b) Fraction of feasible networks in the $f$--$\theta$ plane. The solid line with open circles represents the theoretical boundary as in Fig.~\ref{fig:FSSresult}(b). (c) Probability density function (PDF) of ${N_f^*}/{\ln N_f}$ for $f=0.9$, $0.76$, and $0.5$ for $\theta=-1$ (blue), $\theta=-0.48$ (yellow), and $\theta=0$ (green). 
Each panel shows distributions for $\theta= -1, -0.48,$ and 0. The dashed vertical line marks the marginal feasibility threshold ${N_f^*}/{\ln N_f}=1$, below which networks are classified as non-SF without further FSS analysis.
}
\label{fig:FSSpossibility}
\end{figure*}

To apply the FSS method, we consider BA 
networks of size $N=10^5$ and mean degree $\langle k\rangle=10$, generated as 100 independent realizations. For a given hub-protection level $\theta$, nodes are progressively removed, and the resulting network is analyzed at each point $(f,\theta)$. The scale-freeness is assessed by following the finite-size scaling procedure introduced in previous studies, in which subnetworks of sizes $N/4$, $N/2$, and $3N/4$ are randomly sampled. At each $(f, \theta)$, each subnetwork size is sampled 100 times, and we use the aggregated degree distribution over the 100 samples for a given size $n_f$. Note that $n_f\in\{N_f/4, N_f/2, 3N_f/4, N_f\}$.

Figure~\ref{fig:FSSresult}(b) reports the fraction of realizations that pass the scale-freeness test, defined by simultaneously satisfying the feasibility condition ($N_f^*>\ln N_f$) and a good quality-of-collapse value ($Q\le 3$). Networks with larger $\theta$, corresponding to stronger hub protection, and smaller removal fraction $f$ are more likely to satisfy the FSS criteria, consistent with previous results based on the KL divergence~\cite{lee2022degree}. However, we identify a discrepancy regime in which the KL-divergence criterion classifies networks as SF-like, while the FSS analysis classifies them as non-SF. For reference, the theoretical boundary derived from comparing the KL divergences in Eqs.~(\ref{eq:S_PO}) and~(\ref{eq:S_SF}) is also shown.

Representative FSS results are shown for $(f=0.2,\theta=-1)$ with inconsistent classifications and $(f=0.2,\theta=0)$ with consistent classifications in Figs.~\ref{fig:FSSresult}(c) and~\ref{fig:FSSresult}(d), yielding quality-of-collapse values $Q=18.2$ and $Q=1.25$, and optimal CCDF exponents $\gamma=2.413$ and $\gamma=1.904$, respectively. The original CCDF (insets) displays different slopes at $(f=0.2,\theta=-1)$, whereas they exhibit nearly identical exponents at $(f=0.2,\theta=0)$. Consistently, the scaling collapse is poorer in the former case. Nonetheless, the network at $(f=0.2,\theta=-1)$ remains closer to a static-model SF distribution than to a Poisson distribution. To sum up, the network at the point $(f=0.2, \theta=-1)$ is classified as SF-like by the KL criterion, while failing the FSS test. The combined use of KL divergence and FSS provides complementary insight into the network structure. The results for other types of SF networks are illustrated in Appendix~\ref{app:othersf}.

\subsection{Available degree range and feasibility of FSS}
\label{subsec:feasible}

To better understand what contributes to the classifications in Sec.~\ref{subsec:applyFSS}, we examine the basic statistics required for the FSS analysis. As mentioned above, a sufficient number of nodes must be secured within the scaling range of the degree $k$ to satisfy the feasibility condition.
We first investigate the available degree range between $k_{\rm min}$ and $k_{\rm max}$, where $k_{\rm min}$ is estimated by the KS statistic and $k_{\rm max}$ is taken as the empirical maximum.
The length of the available range, $k_{\rm max}-k_{\rm min}$, shown in Fig.~\ref{fig:FSSpossibility}(a), naturally decreases as the fraction $f$ of removed nodes increases. Moreover, networks with 
smaller $\theta$ exhibit a faster reduction of the degree range,
reflecting the accelerated removal of hub nodes under hub-preferential removal ($\theta<0$). As a result, a narrower range of available degree 
hints at lower feasibility, a smaller value of $N_f^*$.

In Fig.~\ref{fig:FSSpossibility}(b), we plot the fraction of feasible networks out of 100 realizations at each point in the $f$–$\theta$ plane. As described above, infeasible networks are directly classified as non-SF, whereas feasible networks are subjected to the FSS analysis and subsequently classified depending on the value of $Q$. Comparing this with the final classification results in Fig.~\ref{fig:FSSresult}(b), we find that infeasible networks account for the majority of the PO-like classifications by the entropy-based diagnostic, which is consistent with expectation. 

We now examine the inconsistent regimes where the entropy-based diagnostic classifies networks as SF-like, while the scaling-based diagnostic declares them as non-SF. At large $f$ and less negative $\theta$ [red regime near the KL boundary in Fig.~\ref{fig:FSSpossibility}(b)], infeasibility directly leads to the non-SF determination by the FSS method, although the degree distribution can appear more SF-like according to the KL divergence.  This entropy-based classification has limited significance in this regime, given that it is based on a very small number of nodes (of order $\ln N_f$). Meanwhile, at small $f$ and more negative $\theta$ [blue regime near the KL boundary in Fig.~\ref{fig:FSSpossibility}(b)], the networks remain feasible, 
corresponding to a sufficiently wide degree range for scaling, but still fail the FSS test 
due to $Q>3$.

The distributions of $N_f^*/\ln N_f$ are shown in Fig.~\ref{fig:FSSpossibility}(c) for representative values of $f$ and $\theta$.
When a smaller fraction of nodes is removed, most realizations remain feasible and are therefore suitable for the FSS analysis. In contrast, for large $f$, networks under hub-preferential removal become increasingly infeasible for more negative values of $\theta$, whereas most realizations remain feasible for $\theta=0$. This analysis confirms that the fraction of infeasible realizations increases systematically with larger $f$ and more negative $\theta$, indicating that extensive node removal combined with hub-targeted strategies rapidly reduces the effective scaling range.

\subsection{Quality of collapse and estimated exponents}
\label{subsec:Q_gamma}

\begin{figure}[ht]
\includegraphics[width=0.95\columnwidth]{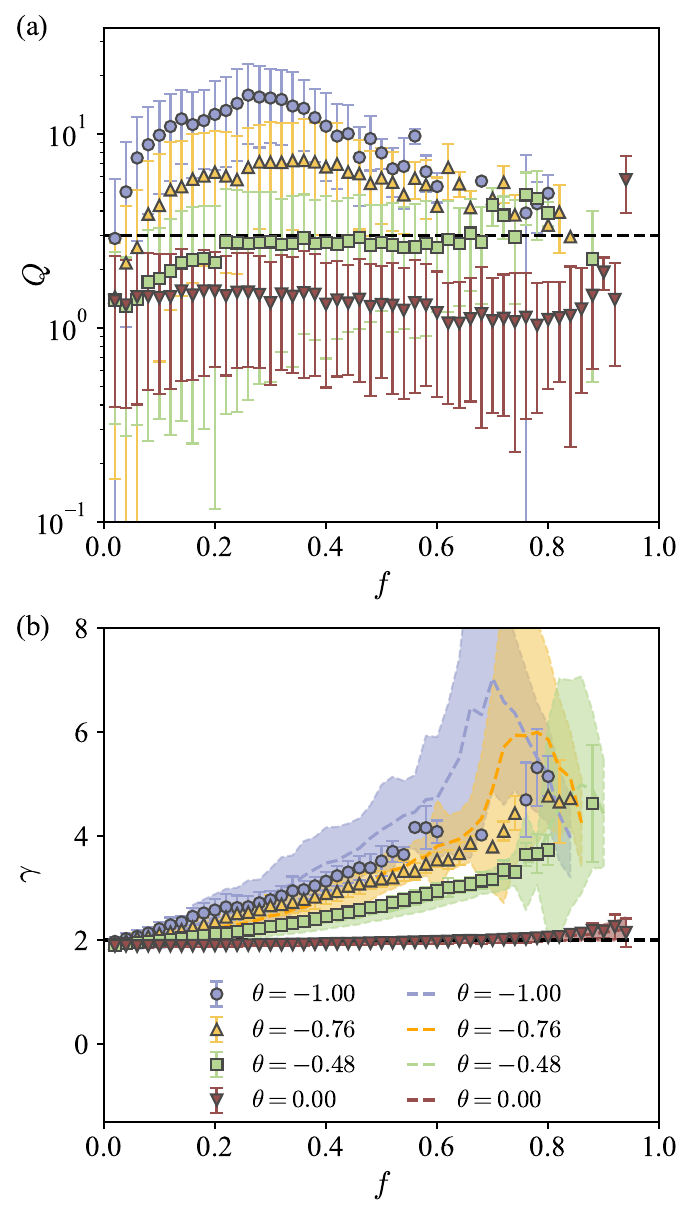}
\caption{Measurements from the FSS analysis as a function of the removal fraction $f$ for $\theta=-1, -0.76, -0.48$, and $0$. (a) Quality of collapse $Q$. Symbols denote simulation results, and lines are guides to the eye. The horizontal black dashed line indicates $Q=3$, which separates SF ($Q\leq 3$) and non-SF ($Q>3$) regimes. (b) CCDF exponent $\gamma$ as a function of $f$
. Filled symbols with error bars show the FSS-based estimates; dashed lines show the MLE-based estimates. Shaded areas indicate the standard deviation of the MLE estimates. The horizontal black dashed line indicates $\gamma=2$. 
}
\label{fig:Q_gamma}
\end{figure}

Based on Fig.~\ref{fig:FSSpossibility}, we perform the FSS analysis by restricting our attention to feasible realizations. Figure~\ref{fig:Q_gamma}(a) shows the quality of collapse $Q$ [Eq.~(\ref{eq:Q})] as a function of the removal fraction $f$. Recall that networks with $Q\leq 3$ are classified as SF, following Ref.~\cite{serafino2021true}. Under hub-protecting removal ($\theta>0$), $Q$ remains close to $1$ even for large $f$, indicating persistent scale-freeness; this behavior is expected, and we do not present these results here. For random removal ($\theta=0$), $Q$ stays near unity at small $f$ but increases sharply at large $f$, signaling a breakdown of scale-freeness. In contrast, under hub-preferential removal ($\theta<0$), $Q$ increases already at an early stage. For $\theta=-0.48$, $Q$ exceeds $3$ only at large $f$, while for $\theta=-0.76$ and $\theta=-1$, it exceeds $3$ already at small $f$ and remains above $Q\gtrsim5$ over a wide range of $f$. 
These behaviors can be understood in terms of the fate of hub nodes: under hub-protecting removal, hubs tend to survive until late stages, leading to the long-lasting persistence of scale-freeness, whereas under hub-preferential removal, hubs are destroyed at early stages, resulting in a premature loss of scale-freeness.

The optimal value of the CCDF exponent $\gamma$ for a given network is determined by minimizing the quality of collapse $Q$. In Fig.~\ref{fig:Q_gamma}(b), we compare this estimate with the exponent $\gamma$ obtained from the maximum-likelihood estimation (MLE)~\cite{clauset2009power}. The two estimates are in good agreement, indicating that the FSS-based procedure provides a reasonable estimation of the scaling exponent. 

For random removal ($\theta=0$) [Fig.~\ref{fig:Q_gamma}(a)], the CCDF exponent remains close to $\gamma=2$, which is the original value for the BA network, and $Q$ is typically below $3$. Only at very large $f$ does $\gamma$ show only a slight increase accompanied by $Q>3$. In contrast, hub-preferential removal leads to an increase of $\gamma$ as the fraction of removed nodes increases. For $\theta=-0.48$, the quality of collapse remains below $Q\lesssim 3$ on average up to $f\lesssim 0.7$ [Fig.~\ref{fig:Q_gamma}(a)], suggesting an apparent persistence of scale-freeness. However, since an SF network is typically associated with the divergence of the second moment $\langle k^2\rangle$ of the degree distribution in the thermodynamic limit, which requires $\gamma<2$, this increase of $\gamma$ implies a progressive weakening of SF heterogeneity, even when a power-law-like tail remains. The estimated values of $\gamma$ also exhibit stronger fluctuations than in the cases $\theta=0$, indicating that the scaling exponent becomes relatively ill-defined and the scale-freeness is less robust. 

This inconsistency persists for more negative values of $\theta$, but takes a different form for $\theta=-1$: although $Q$ decreases after $f\gtrsim 0.4\simeq f^*$ [Fig.~\ref{fig:Q_gamma}(a)], which could be interpreted as a restoration of scale-freeness, the estimated exponent $\gamma$ continues to increase, indicating that the influence of hub nodes is diminishing. Notably, $f\simeq 0.4$ corresponds simultaneously to the crossover point $f^*$ identified by the KL divergence in Sec.~\ref{subsec:node_removal} and to the boundary between feasible and infeasible regimes in Fig.~\ref{fig:FSSpossibility}(b). As a result, the values of $Q$ for $f\gtrsim 0.4$ are obtained from a relatively small number of feasible realizations, making the apparent decrease in $Q$ less reliable as an indicator of scale-freeness. These results show that the behavior of $Q$ alone does not fully characterize the structural state of the network. Therefore, understanding the system requires considering multiple perspectives, rather than relying on the behavior of $Q$ alone.

\subsection{Comparison between FSS and KL-divergence classifications}
\label{subsec:KLdiv}

\begin{figure*}[!th]
\includegraphics[width=\linewidth]{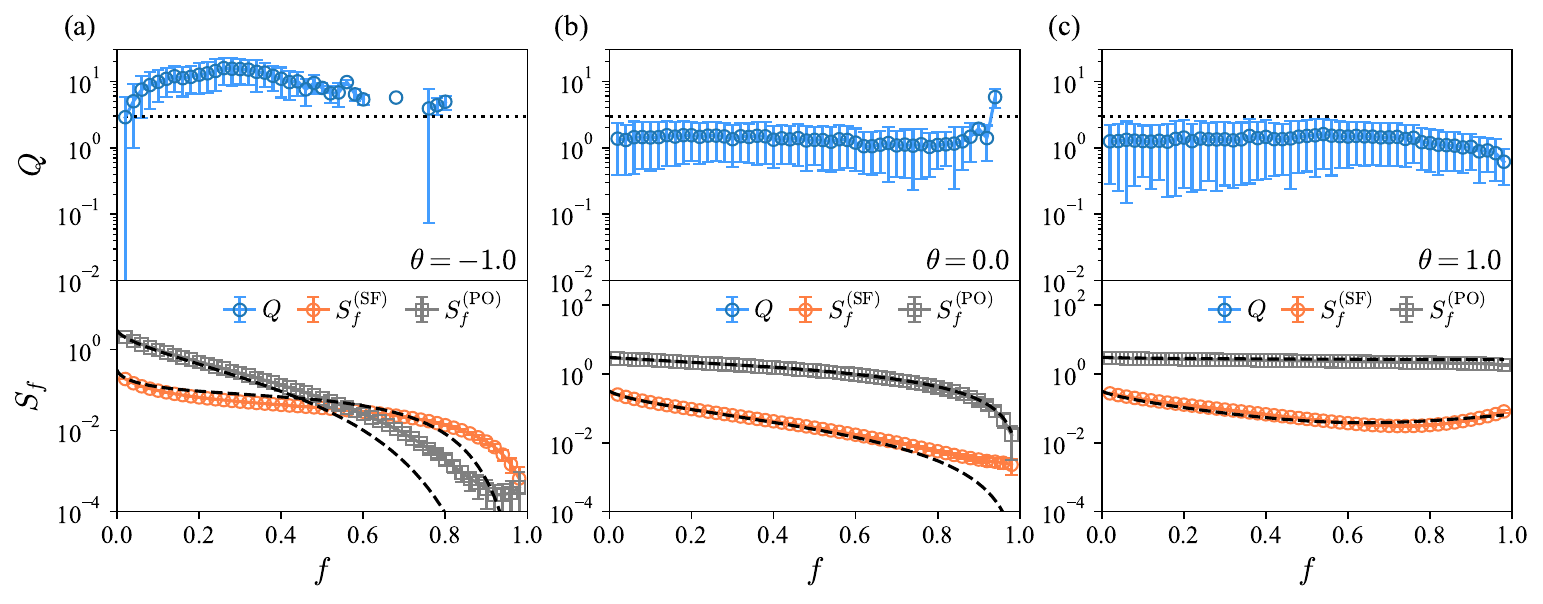}
\caption{The KL divergence $S$ versus the removal fraction $f$ for (a) $\theta=-1$, (b) $\theta=0$, and (c) $\theta=1$. Symbols denote simulation results, and dashed lines show the theoretical predictions from Ref.~\cite{lee2022degree}. The quality of collapse $Q$ (top panels) is replotted from Fig.~\ref{fig:Q_gamma}(a) for comparison with $S_f$ (bottom panels).}
\label{fig:KL_and_Q}
\end{figure*}

We compare the classification of scale-freeness obtained from the FSS analysis with that based on the KL divergence. As summarized in Sec.~\ref{subsec:node_removal}, the boundary separating SF and PO regimes in the KL approach is defined by the condition $S_f^{\rm (PO)}=S_f^{\rm (SF)}$, and the corresponding crossover point is denoted by $f^*$. According to this criterion, networks are classified as SF-like for $f<f^*$, where $S_f^{\rm (PO)}>S_f^{\rm (SF)}$, and as PO-like for $f>f^*$. 

For hub-preferential removal ($\theta=-1$) in Fig.~\ref{fig:KL_and_Q}(a), the KL divergence indicates a crossover from SF-like to PO-like structure at $f^*\simeq 0.4$, whereas the FSS analysis classifies the network as non-SF over the entire range of $f$, as indicated by $Q>3$ [Fig.~\ref{fig:FSSresult}(b)]. In particular, for $f<f^*$, although the degree distribution is closer to the 
static-model SF distribution than to the PO distribution, the system does not exhibit scale-freeness in the scaling-based sense. For $f>f^*$, the KL divergence consistently indicates a PO-like structure, while the FSS analysis identifies the network as non-SF due to the increasing number of infeasible realizations [Fig.~\ref{fig:FSSpossibility}(b)], for which the quality of collapse $Q$ is not defined. In this regime, the absence of a well-defined $Q$ already signals the loss of scale-freeness, whereas the KL divergence further characterizes the resulting structure as being closer to the PO regime.

For $\theta \ge 0$, no crossover between $S_f^{\rm (PO)}$ and $S_f^{\rm (SF)}$ is observed, and the inequality $S_f^{\rm (SF)}<S_f^{\rm (PO)}$ holds over the entire range of $f$ [Figs.~\ref{fig:KL_and_Q}(b) and~\ref{fig:KL_and_Q}(c)], leading the KL-based classification to remain in the SF-like regime [Fig.~\ref{fig:FSSresult}(b)]. Nevertheless, at very large $f$ for $\theta=0$, where the FSS analysis yields $Q>3$, $S_f^{\rm (PO)}$ decreases more rapidly than $S_f^{\rm (SF)}$, even though their ordering is preserved. Because the KL-based classification relies solely on this ordering, it remains insensitive to such differences in the decay rates. This behavior is consistent with the loss of scale-freeness, as evidenced by the increase in $Q$, highlighting a limitation of the KL-based criterion in this regime.

A direct comparison shows that the KL-divergence-based classification and the FSS analysis probe different aspects of network structure. While the KL divergence captures distribution-level similarity to a reference distribution, the FSS analysis tests the robustness of scaling behavior across system sizes. Importantly, the FSS-based assessment itself requires a joint interpretation of multiple quantities, including the quality of collapse $Q$, the optimal exponent $\gamma$, and the feasibility. The combined behavior of these indicators shows that no single quantity is sufficient to characterize scale-freeness under node removal, and a consistent understanding emerges only by considering both the internal consistency of the FSS measures and their relation to the KL-based classification.

\begin{figure*}[t]
\includegraphics[width=1.0\linewidth]{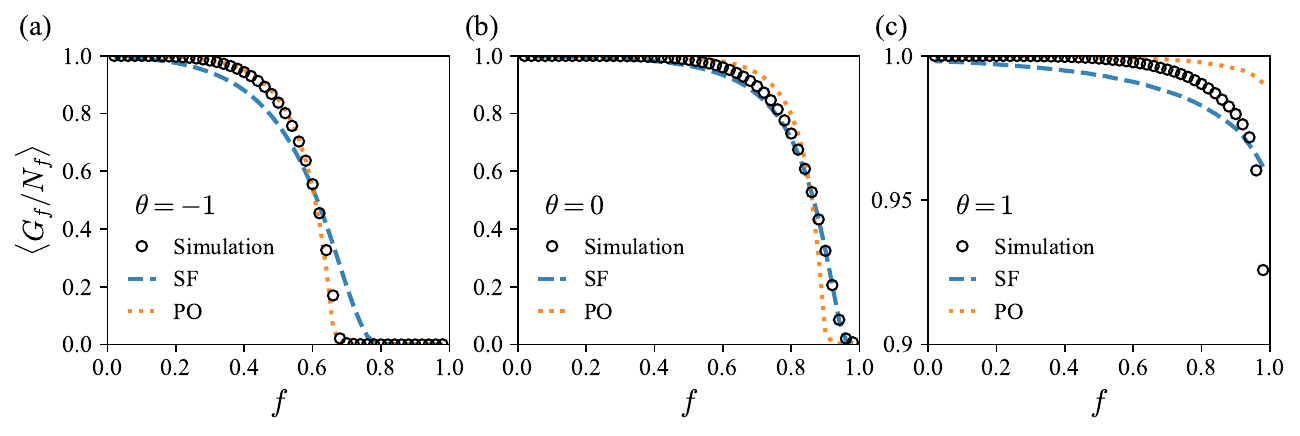}
\caption{
Breakdown behavior of the GCC under degree-dependent node removal. Relative size of the GCC versus the removal fraction $f$ for (a) $\theta=-1$, (b) $\theta=0$, and (c) $\theta=1$. The colored lines are the relative sizes of GCCs of static-model SF networks (blue dashed line) and ER networks (orange dotted line), respectively, with $N_f$ and $m_f$ where $m_f$ is the mean degree at $f$. Symbols are obtained by computing the relative GCC size during the node-removal process, averaged over 100 realizations. 
}
\label{fig:gcc}
\end{figure*}

\subsection{Breakdown behavior of the Giant Connected Component}
\label{subsec:gcc}

While both distribution-based measures (KL divergence and FSS) capture local structural properties, they do not directly reflect global connectivity. In particular, the loss of scale-freeness does not necessarily imply immediate network fragmentation. To assess how removal strategies affect connectivity at the network level, we investigate the GCC, which serves as a proxy for network functionality. We analyze the GCC numerically and examine whether its behavior is consistent with SF-like or PO-like characteristics, in line with the perspective of this study. To provide a baseline for comparison, we consider the GCCs of a static-model SF network and an ER network with the same size $N_f$ and mean degree $m_f$ as the node-removed network. The relative sizes of these GCCs are shown as dashed lines in Fig.~\ref{fig:gcc}. As shown below, the distribution-based classification provides a useful framework for interpreting the observed GCC behavior.

The GCC of the node-removed network for $\theta=-1$ behaves similarly to that of the ER network over a wide range of $f$ [Fig.~\ref{fig:gcc}(a)], consistent with the loss of scale-freeness indicated by the FSS analysis, as hub-preferential removal rapidly destroys the heterogeneous connectivity that distinguishes SF from ER networks. For $\theta=0$, the relative size of GCC is slightly closer to that of the ER network at very low $f$ but mostly follows that of the SF network [Fig.~\ref{fig:gcc}(b)]. For $\theta=1$, the GCC remains well connected, with its relative size close to unity over a wide range of $f$ [Fig.~\ref{fig:gcc}(c)], leaving it difficult to distinguish structural differences based on the GCC in this regime. Nevertheless, we find that the GCC  remains closer to that of the SF network. Overall, these observations show that the distribution-based classification provides a consistent and informative guide to the connectivity behavior captured by the GCC.

\section{Discussion} 
\label{sec:discussion}

In this work, we have investigated how scale-freeness is preserved or lost under node-removal processes by combining the hub-protection framework introduced in Ref.~\cite{lee2022degree} [Eq.~(\ref{eq:qk})] with the finite-size scaling (FSS) analysis proposed in Ref.~\cite{serafino2021true} [Eq.~(\ref{eq:fss_form})]. By applying the FSS framework to node-removed networks, we have shown that scale-freeness is not solely determined by the similarity of degree distributions to reference forms, as quantified by the Kullback-Leibler (KL) divergence [Eqs.~(\ref{eq:S_PO}) and (\ref{eq:S_SF})], but rather by the robustness of scaling behavior across system sizes. In particular, hub-protecting removal ($\theta>0$) tends to preserve scale-freeness over a wide range of removal fractions $f$, whereas hub-preferential removal ($\theta<0$) leads to an early breakdown of scale-freeness, even when the degree distribution remains SF-like. A comparison between KL-based and FSS-based classifications highlights that the two approaches probe distinct aspects of network structure, with the FSS analysis being sensitive to the breakdown of scaling behavior beyond superficial power-law behavior in contracting networks. The overall classification of SF and non-SF regimes is summarized in Fig.~\ref{fig:FSSresult}(b). Consistently, the giant connected component (GCC) analysis provides complementary information on network connectivity that aligns with and helps interpret the trends identified by the distribution-based diagnostics, reflecting that these measured offer a coherent picture across structural scales.
Because each diagnostic is sensitive to a distinct structural scale---local degree statistics, cross-scale invariance, and global connectivity---no single metric can provide a complete picture of structural degradation. This motivates a multi-diagnostic framework as a standard practice in network resilience studies. 

This distinction between KL-based and FSS-based classifications has direct implications for dynamical processes on networks. In the feasible part of the inconsistent regimes, the FSS analysis indicates a loss of scale-freeness while the degree distribution remains SF-like according to the KL divergence, leading to contrasting expectations for epidemic spreading. From the FSS perspective, the loss of scale-invariant organization suggests a suppression of rapid spreading, whereas the persistence of an SF-like degree distribution implies that a non-negligible number of hub nodes may remain, potentially facilitating faster-than-expected propagation. If the latter effect dominates, strategies based solely on entropy-based diagnostics may underestimate the risk of rapid spreading. This observation emphasizes the importance of jointly monitoring both entropy-level and scaling-level indicators when assessing the dynamical vulnerability of networks undergoing structural degradation.


Despite its strengths, the present approach has several limitations. Both the FSS and KL analyses are inherently distribution-based and thus do not capture higher-order interactions or mesoscopic structures, such as community organization or core-periphery structure~\cite{newman2003structure,fortunato2010community,borgatti2000models}, which may also influence network organization. In addition, the FSS analysis relies on random subsampling to construct subnetworks, an unbiased choice that nevertheless does not preserve mesoscopic structures by construction, implying that the effective scaling range and the FSS outcome may depend on the sampling strategy, particularly in correlated networks~\cite{stumpf2005subnets,lee2006statistical}. Nevertheless, by adopting a single, fixed subsampling protocol throughout, we ensure internal consistency across all reported results. These considerations reinforce that scale-freeness in contracting networks cannot be reliably inferred from the form of a single degree distribution alone~\cite{clauset2009power,broido2019scale}, motivating the use of scaling-based diagnostics such as FSS~\cite{serafino2021true}.


In this context, contraction can be viewed as a structural process characterized by path-dependent changes that cannot be inferred solely from static or growing network models~\cite{albert2000error,dorogovtsev2002evolution}. Our results indicate that SF-like degree distributions may persist even when the underlying SF organization, as probed by FSS, is no longer stable. This observation complements earlier work on robustness and vulnerability, particularly in the context of epidemic spreading on networks, where scale-freeness has been shown to play a central role in shaping dynamical responses to damage~\cite{pastor2001epidemic,eguiluz2002epidemic,wang2003epidemic,gomez2007paths,gross2006epidemic}. More broadly, viewing scale-freeness as a property that can be lost during contraction suggests its potential as an indicator of structural integrity in degrading networks~\cite{alstott2009modeling,caeyenberghs2014altered}. Together with global connectivity measures such as the GCC, these results provide a more complete characterization of structural degradation in contracting networks.
\begin{acknowledgments}
We thank Dr. Jung-Ho Kim for providing the code for computing theoretical KL-divergence results. This work was supported by the National Research Foundation (NRF) of Korea through Grant Numbers RS-2026-25488703 (S.-W.S.) and RS-2024-00341317 (M.J.L.), and by a KIAS Individual Grant No. CG079902 (D.-S. L.) at Korea Institute for Advanced Study. We thank APCTP, Pohang, Korea, for their hospitality during the Topical Research Program [APCTP-2025-T04], from which this work greatly benefited. We also acknowledge the hospitality at APCTP, where part of this work was done.
\end{acknowledgments}

\appendix

\section{Other Scale-Free Network Models}
\label{app:othersf}
We have explored the BA model as a representative SF network in the main text. In this section, we extend the analysis to the generalized Barab\'{a}si--Albert (GBA) model and the static model, chosen to have the same CCDF exponent as the BA network. All models are studied for networks of size $N=10^5$ with mean degree $\langle k \rangle = m_{(f=0)} = 10$. Since the GBA model is directed by construction, all analyses are performed using the in-degree $k_{\rm in}$, defined as the number of incoming links to a node.

The GBA model, also known as the initial-attractiveness~\cite{dorogovtsev2000structure} or Price model~\cite{price1976general}, is an extension of the BA model that allows for a variable CCDF exponent $1<\gamma<\infty$, in contrast to the fixed value $\gamma=2$ in the BA model. While the BA model employs the attachment rule $\Pi_i \propto k_i$ for a node $i$, the GBA model generalizes this rule to $\Pi_i \propto k_i + A$, where $A$ is a controllable initial-attractiveness parameter. The CCDF exponent $\gamma$ is related to $A$ as $\gamma = 1 + A/m$, where $m$ denotes the number of stubs introduced by each new node. Because the GBA model is inherently directed and exhibits an SF distribution in the in-degree, our analysis is accordingly performed using in-degree values.

\begin{figure}[t]
\includegraphics[width=1\columnwidth]{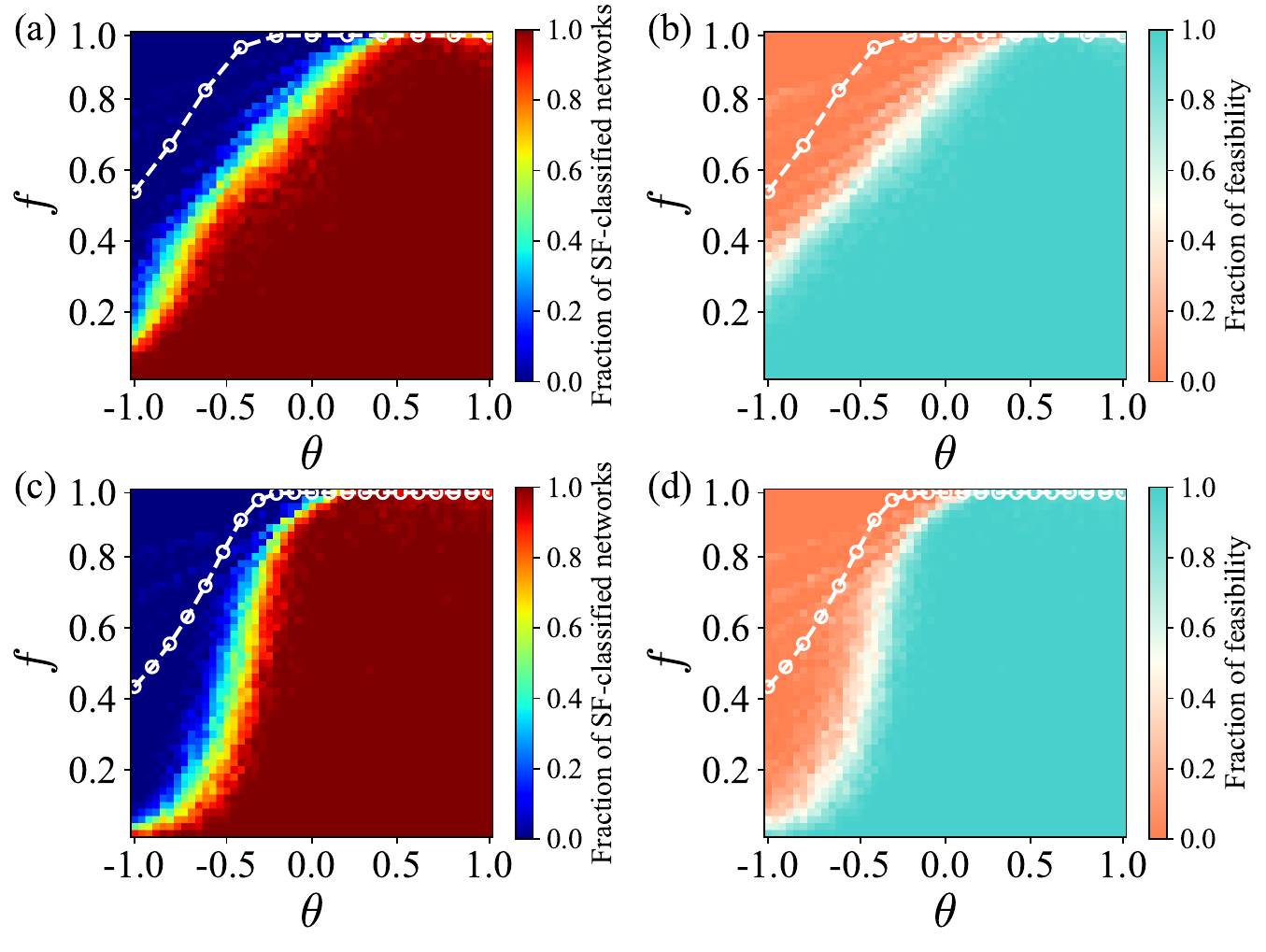}
\caption{Scale-freeness in the $f$--$\theta$ plane for (a, b) the GBA model and (c, d) the static model, both with $\gamma=2$. Results are analogous to those for the BA model shown in Fig. 1(b). (a, c) Classification results. The color bar indicates the fraction of networks classified as SF by the FSS among 100 realizations at each point. (b, d) Feasibility results. The color bar indicates the fraction of feasible realizations. The dashed lines denote the boundary between SF-like (below) and PO-like (above) degree distributions using the KL divergence $S_f^{\rm (PO)}$ and $S_f^{\rm (SF)}$, provided as guides to the eye.
}
\label{fig:app_other_sf_fss}
\end{figure}

The static model~\cite{goh2001universal} is a hidden-variable model for generating SF networks. Each node $i$ is assigned a weight $w_i$, regarded as a hidden variable, given by $w_i \propto i^{-\beta}$, where $\beta \in [0,1)$ controls the weight distribution. Pairs of nodes $i$ and $j$ are connected with a probability proportional to the product of their weights, $w_i w_j$, which yields uncorrelated SF networks in principle. In the limit $\beta \to 0$, the weight heterogeneity vanishes, and the model approaches an ER 
random graph. The weight exponent $\beta$ determines the CCDF exponent as $\gamma = 1/\beta$.

The classification results for the GBA and static models with $\gamma=2$, obtained by setting $A=m=10$ for the GBA model and $\beta=1/2$ for the static model, are displayed in Fig.~\ref{fig:app_other_sf_fss}. Consistent with the BA result in Fig.~\ref{fig:FSSresult}(b), the FSS analysis identifies the SF regime less generously than the KL divergence in both cases [Figs.~\ref{fig:app_other_sf_fss}(a) and~\ref{fig:app_other_sf_fss}(c)]. Comparison with the feasibility fractions in Figs.~\ref{fig:app_other_sf_fss}(b) and~\ref{fig:app_other_sf_fss}(d) shows that infeasible realizations are primarily responsible for the non-SF classification by the FSS analysis.


\providecommand{\noopsort}[1]{}\providecommand{\singleletter}[1]{#1}%

\end{document}